\documentclass[aip,apl,amsmath,amssymb,reprint]{revtex4-1}

\usepackage{graphicx}
\usepackage{dcolumn}
\usepackage{bm}

\usepackage[utf8]{inputenc}
\usepackage[T1]{fontenc}
\usepackage{mathptmx}
\usepackage{etoolbox}

\usepackage{xcolor}

\makeatletter
\def\@email#1#2{%
 \endgroup
 \patchcmd{\titleblock@produce}
  {\frontmatter@RRAPformat}
  {\frontmatter@RRAPformat{\produce@RRAP{*#1\href{mailto:#2}{#2}}}\frontmatter@RRAPformat}
  {}{}
}%
\makeatother
\begin{document}

\preprint{AIP/123-QED}

\title {Fabrication and characterization of Nb / Al-AlN / Nb superconducting tunnel junctions}
\author{A. Pavolotsky}
\author{F. Joint}
\affiliation{Group for Advanced Receiver Development, Department of Space, Earth and Environment, Chalmers University of Technology, 412 96 Göteborg, Sweden}

\author{U. S. Manjunatha}
\affiliation{Materials Research Centre, Indian Institute of Science, Bangalore 560012, India}

\author{D. Meledin}
\author{V. Belitsky}
\affiliation{Group for Advanced Receiver Development, Department of Space, Earth and Environment, Chalmers University of Technology, 412 96 Göteborg, Sweden}

\author{S. Masui}
\affiliation{ALMA project, National Astronomical Observatory of Japan (NAOJ), Mitaka, Tokyo 181-8588, Japan}

\author{T. Kojima}
\altaffiliation[Present address: ]{Quantum Laboratory, Fujitsu Research, Fujitsu Limited, Atsugi, Kanagawa 243-0197, Japan}
\affiliation{ALMA project, National Astronomical Observatory of Japan (NAOJ), Mitaka, Tokyo 181-8588, Japan}

\author{N. Ravishankar}
\affiliation{Materials Research Centre, Indian Institute of Science, Bangalore 560012, India}

\author{V. Desmaris}
\affiliation{Group for Advanced Receiver Development, Department of Space, Earth and Environment, Chalmers University of Technology, 412 96 Göteborg, Sweden}

\email{udupasujit@iisc.ac.in}
\email{joint@chalmers.se}
\email{alexey.pavolotsky@chalmers.se}

\date{\today}

\begin{abstract}
We report a Nb/Al-AlN/Nb superconducting tunnel junction process in which the AlN barrier is formed by plasma nitridation using a compact microwave electron‑cyclotron‑resonance (ECR) nitrogen plasma source integrated into a standard sputter cluster. This enables growth of uniform tunnel barriers across a broad range of specific resistances, with $R_\mathrm{n}A$ down to $\simeq 3~\Omega\,\mu\mathrm{m}^2$. Junctions maintain excellent quality, exhibiting $R_\mathrm{j}/R_\mathrm{n}\ge 25$ at the highest barrier transparencies. We characterize resistivity, specific capacitance, and the evolution of junction parameters under room‑temperature aging and thermal annealing. A consistent calibration of the junction specific capacitance $C_s$ versus $R_\mathrm{n}A$ is established and independently validated by the performance of demonstrator SIS mixers designed using the extracted $C_s$.
\end{abstract}

\maketitle

\section{\label{introduction}Introduction}
Superconductor-Insulator-Superconductor (SIS) tunnel junctions are the key nonlinear element in a wide range of superconducting technologies, including quantum-limited heterodyne mixers for millimeter and submillimeter radioastronomy, SQUIDs, voltage standards, digital SFQ circuits, and qubits. The standard niobium-based SIS junction uses an ultrathin aluminium oxide, AlOx, barrier developed in the early 1980s \cite{gurvitch1981,gurvitch1983}, which enabled highly uniform and reproducible Nb/Al-AlOx/Nb trilayers for both analog and digital superconducting electronics.

Despite these advantages, AlOx-barrier junctions face intrinsic limitations when pushed toward higher current densities and operating frequencies. The dependencies of junction superconducting critical current $J_C$ and  specific resistance $R_{n}A$ (where $R_n$ and $A$ are junction's normal state resistance and area, respectively) on oxidation exposure become increasingly steep for highly transparent SIS junctions needed for operation at higher frequencies. The thinner the barriers, the more they are prone to thickness fluctuations and structural defects, e.g., oxygen-vacancy clusters and microscopic shorts. These inhomogeneities lead to a degradation of uniformity and enhanced subgap conductance. Moreover, AlOx-barrier SIS junctions with the high barrier transparency needed for higher-frequency operation are characterized by specific capacitance values that are problematically high for device operation at terahertz frequencies.

To address these challenges, alternative barrier materials have been investigated. Aluminium Nitride (AlN) offers a promising route, providing barriers with lower specific capacitance and higher transparency.
Several fabrication approaches have been developed for Nb/Al-AlN/Nb junctions, including plasma nitridation cathode- \cite{shiota1992, dolata1995,kleinsasser1995} or anode-biased \cite{bumble2002,iosad2002,iosad2003,dmitriev2003,akaike2013}, inductively coupled RF plasma \cite{Lichtenberger2009}, microwave electron-cyclotron resonance (ECR) \cite{endo2009}, ion-beam assisted nitridation \cite{kaul2005}, and direct AlN deposition \cite{zhang2024}. Among these different methods, the ones that maximize nitrogen radical density while keeping ion bombardment at a low level are particularly favorable, since they enable fabrication of highly transparent barriers with low subgap leakage.

From the perspective of SIS mixers, one of the most critical performance metrics is the ratio $R_j/R_n$, where $R_j$ is the subgap resistance and $R_n$ is the normal state resistance. The subgap leakage current directly increases mixer noise \cite{feldman1991,ke1993} and those reported junctions with specific resistance $R_{n}A\leq 10\,\Omega\cdot\mu m^2$ typically show $R_j / R_n \leq 15$. Achieving higher $R_j / R_n$ ratios and high transparency requires carefully tuned AlN barriers with optimal plasma conditions.

In this work, we present the development of Nb/Al-AlN/Nb SIS junctions fabricated using a compact microwave ECR plasma source (Aura-Wave \cite{latrasse2016}), which provides high radical density and is readily integrated into a sputtering system. We investigate the junction resistivity and specific capacitance, and we validate the capacitance calibration through SIS mixer measurements. We demonstrate that AlN barriers fabricated under optimized plasma conditions can simultaneously achieve high transparency and suppressed subgap leakage, improving the previously reported $R_j/R_n$. Our results open perspectives for low-noise SIS mixers and advanced quantum circuits. 

\section{\label{Experimental}Experimental}

\subsection{\label{exp:fab}Fabrication process}

The starting point for the $Nb/Al-AlN/Nb$ junction fabricating process reported here was our well-established process for Nb/Al-AlOx/Nb junctions \cite{pavolotsky2011}, which for years has been employed to fabricate high-quality junctions and mixers made from them. The only change to that baseline was to replace thermal oxidation by plasma nitridation of the Al layer. 

The $Nb/Al$--$AlN/Nb$ trilayer was grown in a single vacuum run by means of dc magnetron sputtering. The bottom Nb layer of the trilayer, 200 nm thick, was deposited at a rate of 0.9 nm/s, followed by about 7 nm of Al deposited at 0.3 nm/s. All layers were deposited on the substrate mounted to the massive copper holder placed over the water-cooled deposition station.
The fresh Al surface was exposed to plasma nitridation, as discussed in the paragraph below, at room temperature. Subsequently, a 100 nm-thick Nb layer was deposited under the same conditions as the bottom Nb layer. The base electrode pattern was etched with reactive ion etching (RIE) through the $Nb/Al–AlN/Nb$ trilayer, using $\mathrm{CF}_4 + \mathrm{O}_2$ for Nb and $\mathrm{Cl}_2$ for the Al–AlN layers, respectively. The junction pattern was deﬁned by the same RIE process with a stop at the $Al–AlN$ layer, followed by a deposition of a 150 - 250 nm thick $\mathrm{SiO}_2$ layer by means of reactive rf magnetron sputtering. A 300 - 400 nm thick Nb wiring layer was deposited by dc sputtering and further patterned by RIE.

For nitridation, we used a compact Aura-Wave ECR plasma source \cite{latrasse2016}, installed in place of one the magnetron positions in our sputtering cluster tool \cite{pavolotsky2011}. To protect the plasma source from side-deposition by other sputter cathodes, it was enclosed into a grounded cylindrical sputtering shield. The Al exposure time to the nitridation plasma was controlled by movement of the deposition shutter. During nitridation, the substrate could be left electrically grounded or biased by a few volts; only grounded substrate results are discussed in this paper.

The ECR plasma density proved to be sufficiently high that, to reach practically relevant values of $R_n A$, we operated the source at the lowest microwave power compatible with stable operation, i.e. 12\ldots 15\,W for all data reported here. For the same reason, nitrogen was diluted with argon. We did not throttle the pumping; the chamber pressure was set solely by the nitrogen and argon flow rates, 1 - 4 sccm and 12 - 16 sccm, respectively. The total pressure during nitridation was $(1-2)\cdot 10^{-3}$\,mbar.

\subsection{\label{exp:char}Junction characterization}

High-resolution annular dark field scanning transmission electron microscopy (ADF-STEM) imaging of nitridized Al surface was performed on a Cs probe corrected ThermoFisher® Titan® Themis™ G2 microscope operating at 300 kV. In conjunction with ADF-STEM imaging, electron energy loss spectroscopy (EELS) analysis was performed using the same TITAN G2 microscope equipped with a Gatan Quantum 965 imaging filter, operating at 300 kV. 

Test samples contained circular SIS junctions with nominal diameters of 1.6, 2.0, 2.4 and \,$\mu\mathrm{m}$. In this work, for junction characterization, we measured their current-voltage characteristics (IVC) at $\approx 3$\,K in the closed-cycle cryostat. Analyzing the recorded IVCs, we extracted the values of junction normal resistance, $R_n$, and the subgap resistance $R_j$.
We estimated $R_n$ as an average resistance measured in the range of $\sim 1.3\cdot V_g\ldots V_{max}$, where $V_{max}$ is the maximum bias voltage value during recording of IVCs  (in this work, typically 6\,mV) and $V_g$, the superconducting gap voltage, taken as the voltage at the maximum slope of the superconducting gap branch of the IVC; $R_j$ was taken as the slope of the line drawn from the I-V origin and tangential to the subgap branch of the IVC.

In the course of the fabricating process, the sizes of SIS junctions are subject to an offset $t$ associated with lithography and etching. To extract the specific resistance $R_nA$ of a given batch of fabricated junctions, we fitted the dependence of the measured $R_n$ of the test junctions on their nominal area $A_{nom}$ to the analytical expression (\ref{eq:RnA}) below having RnA and offset $t$ as fitting parameters \cite{lopez2025}:

\begin{equation}
    R_n = \frac{R_\mathrm{n}A}{\pi\left(\sqrt{\frac{A_{nom}}{\pi}}-t\right)^2}\label{eq:RnA}
\end{equation}

We tracked changes in the normal resistance $R_n$ and the subgap resistance $R_j$ under room-temperature aging and stepwise annealing. Chips were baked on a hotplate in air for 1\,hour at each set point from $130^{\circ}\mathrm{C}$ to $200^{\circ}\mathrm{C}$ in $10^{\circ}\mathrm{C}$ increments. After each step, I-V characteristics were measured to extract updated $R_n$ and $R_j$.

\subsection{Specific capacitance measurement}
We determined the junction specific capacitance $C_s$ following two approaches:

\emph{(i) Cryogenic S‑parameter extraction.} Following Ref.~\cite{aghdam2015meas}, junctions were connected with on-chip microstrip lines to coaxial connectors 
and characterized in a 4K closed-cycle cryostat using S-parameter measurements~\cite{rashid2014IFhybrid}. $C_s$ was extracted by fitting the measured $S_{11}$/$S_{21}$ to an equivalent circuit in which the SIS element is represented by its small-signal complex admittance and a parallel capacitance.  Because long stainless-steel cryo cables are required between the device under test at 4\,K and the room temperature VNA, we de-embedded cable loss/phase.

\emph{(ii) On‑wafer cryogenic probing.} To mitigate cable and temperature gradient uncertainties, we also measured $C_s$ directly on the wafer in a cryogenic probe station, performing a Short-Open-Load (SOL) one-port S-parameter calibration referenced to the probe tip using calibration standards on an adjacent substrate. This approach provides near‑direct access to the device pads and reduces the residual systematic error \cite{kojima2017on-wafer,kojima2019contribution}.

\subsection{\label{exp:mixer}Mixer measurements}

To validate both junction performance and the $C_s$ calibration, we characterized a demonstrator SIS mixer using the standard Y‑factor method \cite{pozar2021microwave}. With $Y=P_\mathrm{Hot}/P_\mathrm{Cold}$, the double‑sideband (DSB) receiver noise temperature is
\begin{equation}
T_\mathrm{rec}=\frac{T_\mathrm{H}-Y\,T_\mathrm{C}}{Y-1}
\end{equation}
Hot/cold loads of known physical temperatures $T_H$ and $T_C$ were used, and measurements were taken across the IF band for representative LO frequencies. The quoted $T_\mathrm{rec}$ values include optical losses in the laboratory setup; based on measured transmission we estimate an added 12–16~K from 275 to 375~GHz. No correction for this contribution was applied. IF ripples arise from reflections between the mixer and the cryogenic LNA (4–16~GHz) in the absence of an IF isolator; this choice enabled probing the full IF tuning range at the cost of stronger standing‑wave structure.

\section{\label{Results}Results and discussion}

\subsection{\label{res:HRTEM}Imaging and analysis of grown AlN tunnel barrier}

\begin{figure}
\centering
\includegraphics[width=0.75\columnwidth]{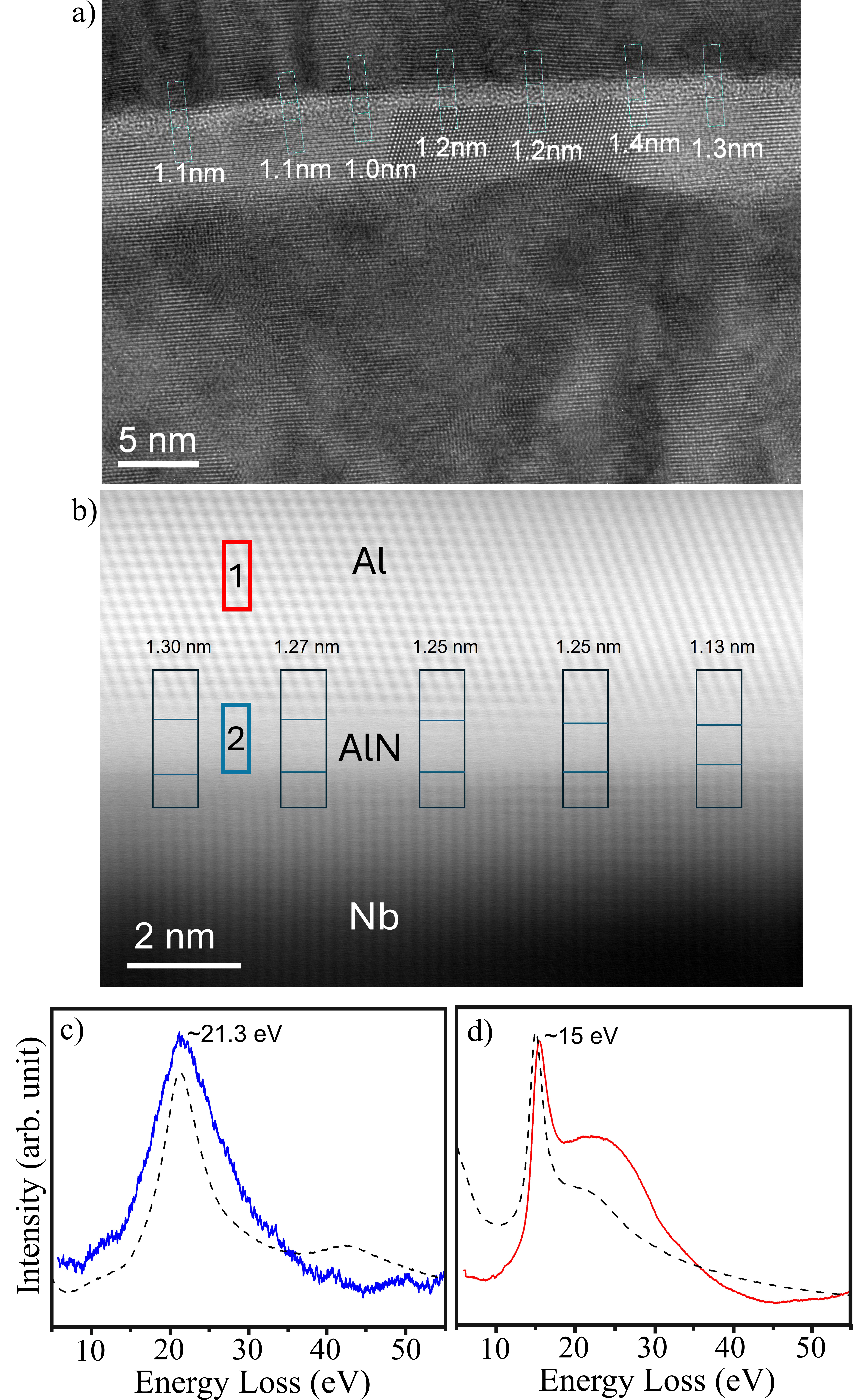}
\caption{\label{fig:HRTEM} \textit{(a)} ADF-STEM micrograph of Nb/Al-AlOx/Nb tunneling structure characterized with $R_\mathrm{n}A$\,$\approx 30\,\Omega\cdot\mu m^2$ reproduced from Ref.~\cite{aghdam2016dependence} compared with \textit{(b)} of Nb/Al-AlN/Nb tunneling structure with $R_\mathrm{n}A$\,$\approx 5\,\Omega\cdot\mu m^2$ . \textit{(c)} low loss plasmon peak of AlN as recorded in the region 2 corresponding to the tunnel barrier layer seen on the micrograph \textit{(b)} - \textit{blue} curve, as compared to the database plasmon peak spectrum from Ref.~\cite{serin1998eels} - \textit{dotted} curve.\textit{(d)} low loss plasmon peak of Al as recorded in the region 1 of the micrograph \textit{(b)} - \textit{red} curve, as compared to the standard EELS plasmon spectra of Al taken from EELS atlas - \textit{dotted} curve.}

\end{figure}

The structure and composition of the fabricated AlN tunnel barrier were investigated using high-resolution ADF-STEM in conjunction with low-loss electron energy loss spectroscopy (EELS). Figure~\ref{fig:HRTEM}b displays a representative ADF-STEM micrograph of a Nb/Al-AlN/Nb tunnel junction, demonstrating $R_\mathrm{n}A$ $\approx 5\,\Omega\cdot\mu m^2$. Based on an intensity line profile across the barrier, as depicted in the micrograph, the AlN thickness is estimated to be in the range of $1.2$–$1.5$\,nm. This value is significantly larger than AlOx tunnel barrier thickness in the Nb/Al-AlOx/Nb junction batch with $R_\mathrm{n}A \approx 30~\Omega\,\mu\mathrm{m}^2$ (Fig.~\ref{fig:HRTEM}a, reproduced from Ref.~\cite{aghdam2016dependence}). 
To chemically verify the composition of the barrier, low-loss EELS spectra were acquired from two distinct locations indicated in the micrograph: region 1 (red) and region 2 (blue).
A clear energy shift of approximately $6.3$\,eV was observed in the bulk plasmon peak between these two regions.
The peak corresponding to region 1 (Fig.~\ref{fig:HRTEM}d), located at $\sim$15\,eV, aligns with the accepted value for metallic aluminium, whereas the peak for region 2 (Fig.~\ref{fig:HRTEM}c) at $\sim$21.3\,eV is in excellent agreement with the reference value for aluminium Nitride from the EELS database~\cite{serin1998eels}. Importantly, within the aluminim Nitride layer, the observed plasmon peak was consistently at $\sim$21.3\,eV and no shift or broadening of the peak was observed, hence excluding the presence of graded $AlN_x$ phases. Altogether, this observation unequivocally confirms uniform formation of the intended stoichiometric aluminium Nitride layer.

\subsection{\label{res:ivc}Nb/Al-AlN/Nb junctions' resistivity}

\begin{figure}
\centering
\includegraphics[width=0.75\columnwidth]{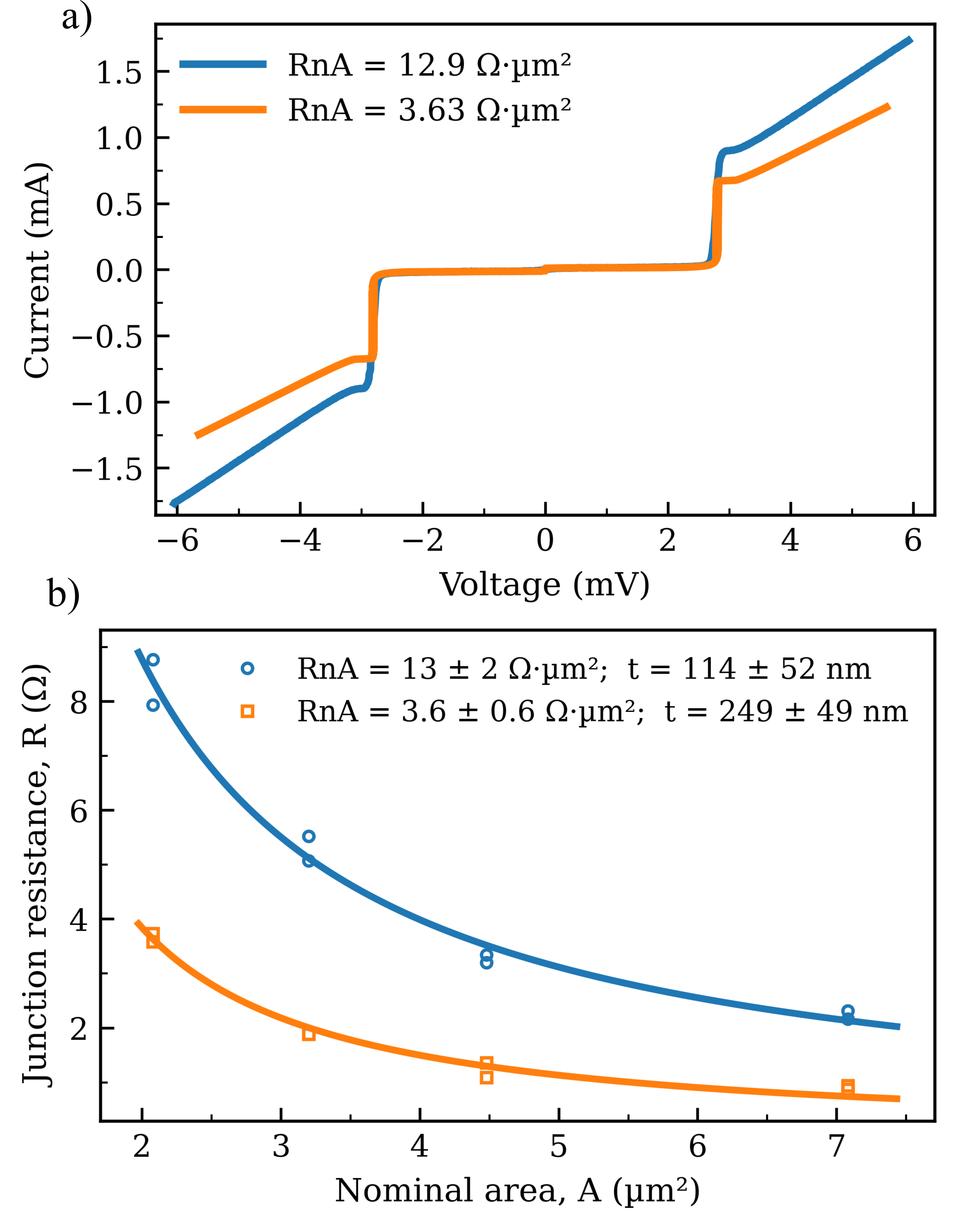}
\caption{\label{fig:IVCs}Current-voltage characteristics of Nb/Al-AlN/Nb SIS junctions. \textit{(a)} Junction with nominal area $0.8~\mu\mathrm{m}^2$) and $4~\mu\mathrm{m}^2$ and extracted $R_\mathrm{n}A \approx 3.6~\Omega\,\mu\mathrm{m}^2$ and $R_\mathrm{n}A \approx 13~\Omega\,\mu\mathrm{m}^2$, respectively. The $R_\mathrm{n}A$ values are determined following Eq.,(\ref{eq:RnA}). \textit{(b)} shows the corresponding wafer-level statistics used to extract the $R_\mathrm{n}A$ values for the junctions displayed in \textit{(a)}.}
\end{figure}

Junctions with specific resistance down to as low as $R_\mathrm{n}A\approx 3~\Omega\,\mu\mathrm{m}^2$ were fabricated while retaining excellent subgap performance, with $R_\mathrm{j}/R_\mathrm{n}\ge 25$ even at the highest tunnel barrier transparencies (Fig.~\ref{fig:IVCs}). 
Although we did not target $R_\mathrm{n}A<3~\Omega\,\mu\mathrm{m}^2$, no trend suggested imminent degradation as transparency increased toward this limit. The observed back-bending of the gap branch of the $R_\mathrm{n}A\approx 3~\Omega\,\mu\mathrm{m}^2$ junction's IVC is likely due to local heating of the junction by the measurement current.

\subsection{\label{res:thermal}Aging and annealing performance}

To quantify thermal robustness and the suitability of AlN‑barrier SIS junctions for microfabrication flows that include modest post‑deposition heating, we tracked changes in the normal resistance $R_n$ and the quality factor $R_j/R_n$ after room‑temperature aging and during a stepwise anneal. Devices from two $R_nA$ ranges ($\sim 15$ and $\sim 120~\Omega\cdot\mu\mathrm{m}^2$) were baked on a hotplate in air for 1\,h at each setpoint from $130^{\circ}\mathrm{C}$ to $200^{\circ}\mathrm{C}$ in $10^{\circ}\mathrm{C}$ increments. After each step, we recorded IVCs at $\sim 3$\,K and re‑extracted $R_n$ and $R_j/R_n$.

\begin{figure}
\centering
\includegraphics[width=0.75\columnwidth]{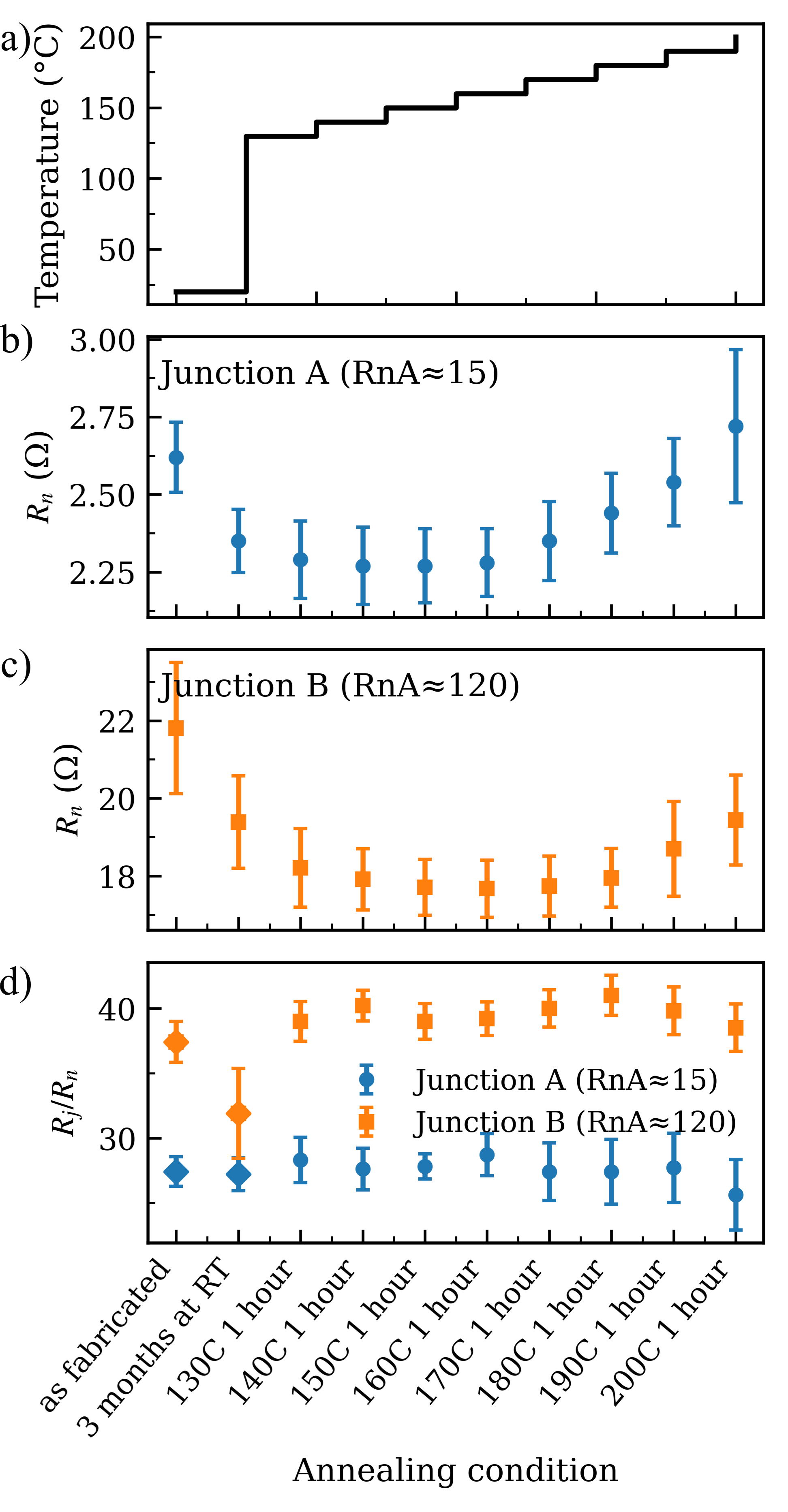}
\caption{\label{fig:annealing}Aging and annealing behavior of Nb/Al-AlN/Nb junctions: \textit{(a)} Thermal profile of the annealing process. \textit{(b)} and  \textit{(c)} Evolution of normal resistance $R_n$ for junctions A and B, respectively. \textit{(d)} junctions quality factor $R_j /R_n$. Results are shown for junctions batches with $\ R_nA\sim 15\,\Omega\cdot\mu m^2$ and $\ R_nA\sim 120\,\Omega\cdot\mu m^2$.}
\end{figure}

Figure~\ref{fig:annealing} summarizes the results. Across both $R_nA$ ranges, we observe (i) a modest decrease in $R_n$ of typically $<20\%$ upon annealing to $160$–$170^{\circ}\mathrm{C}$, followed by (ii) a minor increase as the temperature is raised to $200^{\circ}\mathrm{C}$.The absence of an increase in sub‑gap leakage even in the most transparent junctions is indicating that annealing does not create barrier defects or activate additional conduction channels.

A simple physical picture consistent with these trends is the competition between (a) ordering/densification of the ultrathin AlN through stress-induced flow of vacancies in the adjacent Al layer, as shown in Ref.~\cite{pavolotsky2011}, which effectively thins the tunnel barrier and lowers $R_n$ at the lower set points, and (b) a minor increase of the junction normal resistance due to accommodation of molecules absorbed at the AlN/Nb interface from the residual vacuum atmosphere (mostly, water vapour). Compared with our earlier AlO$_x$ junctions fabricated in the same toolset \cite{pavolotsky2011}, AlN barriers require higher temperature for a comparable magnitude of the $R_n$ change, suggesting that the as‑grown AlN is more structurally ordered than room‑temperature AlO$_x$ and therefore less susceptible to low‑temperature relaxation. The shorter growth exposure for AlN (microwave nitridation $\sim$\,1\,min) relative to AlO$_x$ oxidation ($\sim$\,15\,min) also reduces the opportunity for uptake of residual adsorbates that could react during anneals and permanently thicken the barrier.

The demonstrated temperature stability of Nb/Al-AlN/Nb junctions opens possibility for various fabrication and packaging processes (e.g., resist baking, deposition processes like PECVD and PEALD, chip bonding, gluing etc.), and confirms the suitability of these junctions for use in devices operating for many years without degradation of their properties.

\subsection{\label{res:capac}Nb/Al-AlN/Nb junctions' specific capacitance}

\begin{figure}
\centering
\includegraphics[width=0.75\columnwidth]{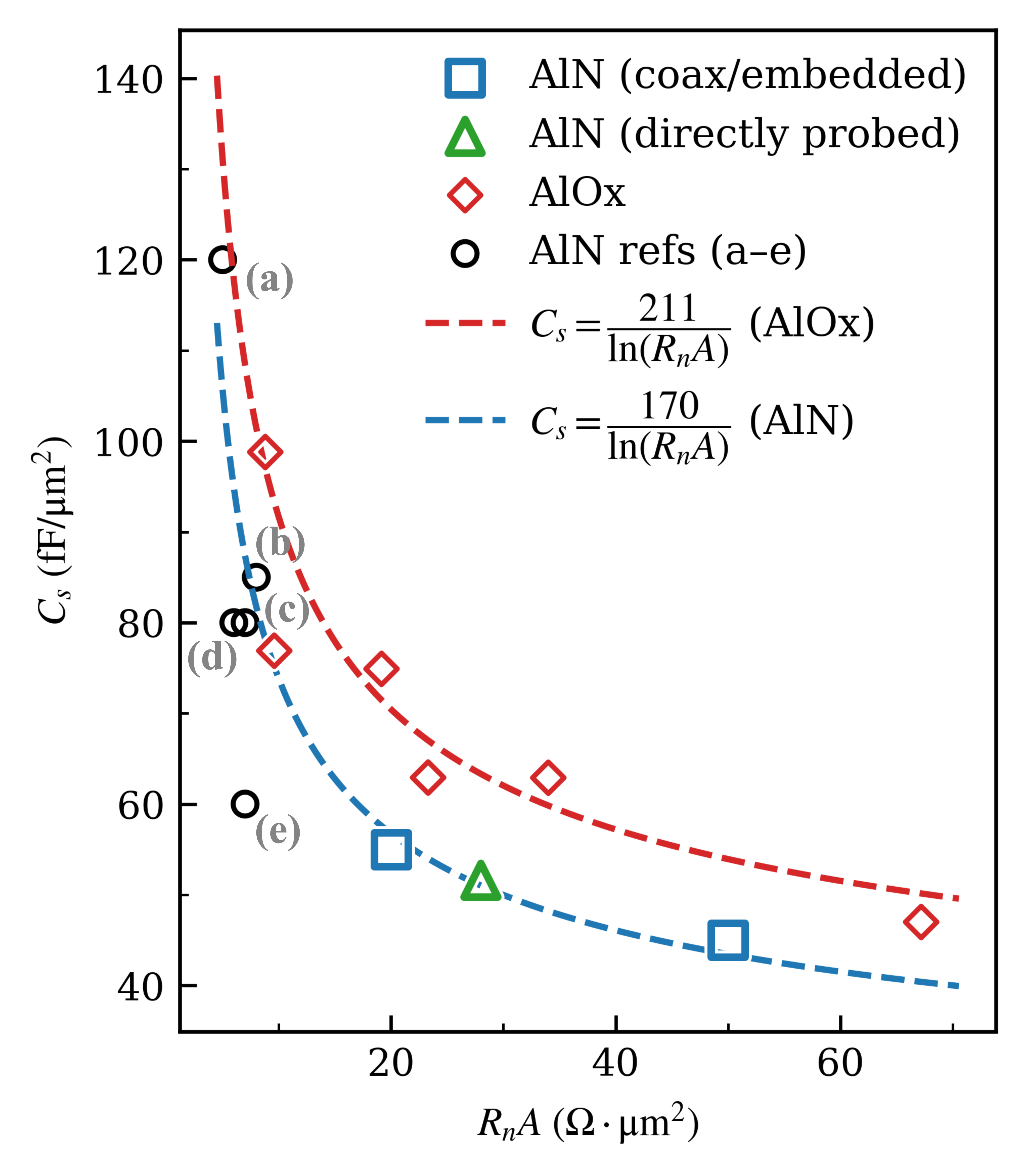}
\caption{\label{fig:Cs} Specific capacitance $C_S$ of Nb/Al-AlN/Nb junctions as a function of $R_\mathrm{n}A$. Results from direct probe measurements are shown as green triangle and values extracted from S-parameter measurements as light blue squares. For comparison, capacitance data reported by other groups are included (dark blue circles): \textit{(a)} Ref.~\cite{kojima2018}, \textit{(b)} Ref~\cite{kawamura2000}, \textit{(c)} Ref.~\cite{kooi2020}, \textit{(d)} Ref.~\cite{khudchenko2015}, \textit{(e)} Ref.~\cite{lodewijk2009}. Data for Nb/Al-AlOX/Nb junctions are plotted as red diamonds \cite{aghdam2017Cs}. The capacitance dependence is approximated by a semi-empirical relation~\cite{belitsky1993Cs_vs_RnA} $C_s = a / \ln(R_\mathrm{n}A)$  with $a = 211$ for AlOx-barriers junctions \cite{aghdam2017Cs} and $a = 170$ for the present AlN-barrier junctions.
}
\end{figure}

We determined the specific capacitance $C_s$ using the two complementary methods, S-parameter measurements and on-wafer probing, and then cross-validated the results. Figure~\ref{fig:Cs} compiles both data sets together with representative literature values. For the three $R_nA$ points measured in this work, the two methods agreed within $\sim\!10$–$15\%$ after de‑embedding. 
Fitting the combined data to the semi‑empirical relation \cite{belitsky1993Cs_vs_RnA}

\begin{equation}
C_s = \frac{a}{\ln(R_nA)}\ ,
\label{eq:Cs_vs_RnA}
\end{equation}

with $R_nA$ in $\Omega\cdot\mu\mathrm{m}^2$ and $C_s$ in fF/$\mu\mathrm{m}^2$, yields $a=170$ for AlN‑barrier junctions (best fit across our $R_nA$ span) versus $a=211$ previously established for our AlO$_x$ process. The residuals of the fit on our data are within the combined experimental uncertainty. As a result, at a given $R_nA$ the AlN barrier provides $\approx 25\%$ lower specific capacitance than AlO$_x$ (red diamond in Fig.~\ref{fig:Cs}), consistent with most prior reports over a comparable tunnel barrier transparency range.

\emph{Physical interpretation.} For a fixed transparency (i.e., fixed $R_nA$), a lower effective tunneling barrier height and/or electron effective mass in AlN relative to AlO$_x$ implies a slightly thicker equivalent barrier, which directly reduces $C_s$. The HRTEM/EELS in Sec.~\ref{res:HRTEM} corroborates that our AlN barriers are indeed thicker than AlO$_x$ barriers in devices of similar $R_nA$, supporting the capacitance trend.

\emph{Implications for circuit design.} The $\sim 25\%$ reduction in $C_s$ eases the reactive matching of high‑transparency SIS junctions by pushing the intrinsic RC pole to higher frequency at a fixed $R_nA$. From a mixer perspective, this enables a wider RF/IF band of operation achieved by limiting the reactive part of the junctions' impedances.

\begin{figure}
\centering
\includegraphics[width=0.75\columnwidth]{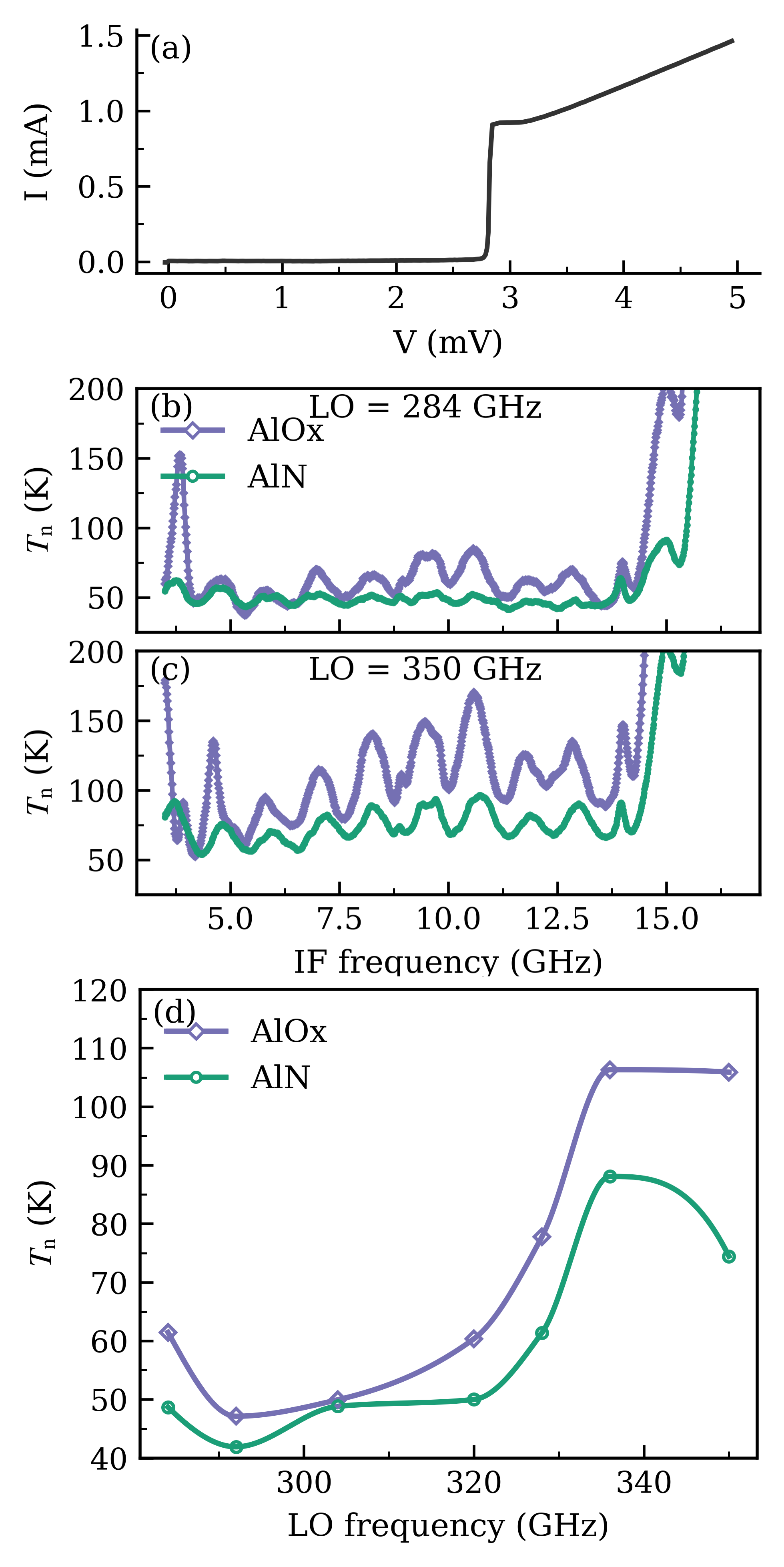}
\caption{\label{fig:mixer} Performance of the demonstrator Nb/Al–AlN/Nb SIS mixer: \textit{(a)} Current–voltage characteristic of the mixer junction. 
\textit{(b)}–\textit{(c)} DSB receiver noise temperature $T_\mathrm{n}$ versus IF frequency for LO = 284 GHz and 350 GHz, respectively. 
\textit{(d)} $T_\mathrm{n}$ versus LO frequency, averaged over the IF band 3.5–14 GHz. 
Results are shown for both AlOx- and AlN-barrier junctions.}
\end{figure}

We validated the calibration by re‑using an established 275–370\,GHz AlO$_x$‑based chip layout \cite{risacher2006_APEX2A} and adjusting only the SiO$_2$ thicknesses.
The AlN‑based demonstrator exhibited equal or slightly wider RF/IF tuning and a comparable or lower DSB noise temperature (Fig.~\ref{fig:mixer}), consistent with the $C_s$ reduction and the high $R_j/R_n$ observed in our IVCs.

\begin{figure}
\centering
\includegraphics[width=0.75\columnwidth]{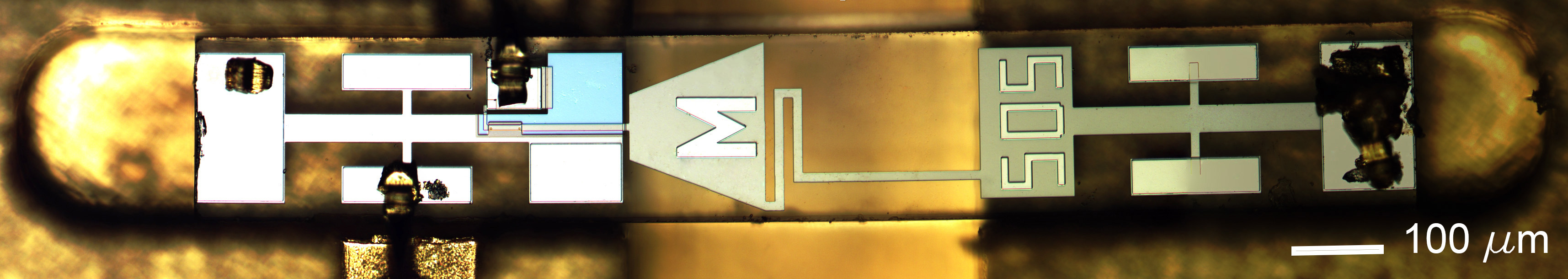}
\\a)\vspace{0.2cm}\\
\includegraphics[width=0.75\columnwidth]{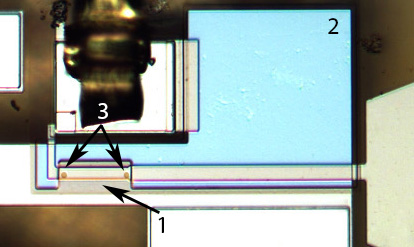}
\\b)\vspace{0.2cm}\\
\caption{\label{fig:chip} \textit{(a)} Mixer chip view. \textit{(b)} Closeup view at the part of the chip containing SIS junctions: 1 -- $SiO_2$ layer between choke and the microstrip line around the twin junctions; 2 -- $SiO_2$ layer between choke and the microstrip line of the impedance transformer; 3 -- twin SIS junctions.}
\end{figure}

The measured noise performance of the demonstrator mixer made with Nb/Al-AlN/Nb superconducting tunnel junctions compared to those with Nb/Al-AlOx/Nb junctions is shown in Fig.~\ref{fig:mixer}. The measured noise temperature includes the contribution of the optical losses in the laboratory mixer measurement setup. That is due to losses in the plastic lens placed at the 150K thermal shield and slight spillover loss at the opening of the cryostat's vacuum window. We estimate the added noise temperature to be approximately 12~K at about 275~~GHz and up to roughly 16~K at 375~GHz. The noise temperatures shown in Fig.~\ref{fig:mixer} are not corrected for this additional noise contribution. 

The variations of the noise temperature seen on the plots against intermediate frequency (IF), Fig.~\ref{fig:mixer}b,c, of the mixer are due to the reflections between the mixer and cryogenic low-noise amplifier (LNA). Compared to the SIS mixers operated at the APEX telescope in the SHeFI receiver \cite{risacher2006_APEX2A,vassilev2008SHeFI}, we used a 4–16 GHz LNA instead of a 4–8 GHz LNA at the telescope; we did not use 4-8 GHz IF isolator. Also, compared to SIS mixers operated at the APEX telescope in the SEPIA 345 receiver \cite{belitsky2018sepia,meledin2022sepia345}, we did not use 4-12 GHz IF isolator. This allowed probing of the full range of IF tuning of the mixers, but at the cost of higher reflections in the IF signal path.

Comparing the noise temperature plots against both the IF and the local oscillator (LO) frequency, we can see that the demonstrator mixer with AlN barrier junctions has the same or slightly wider IF/RF tuning, thus proving the correctness of the calibration of $C_s$ vs $R_\mathrm{n}A$. The somewhat lower noise temperature shown by the mixer with AlN-barrier junctions is likely due to superior Nb/Al-AlN/Nb junctions quality, $R_j/R_n > 30$, as seen on the Fig.~\ref{fig:mixer}a. 

\section{\label{Conclusion}Conclusion}

We have demonstrated a reliable process for Nb/Al–AlN/Nb SIS junctions in which a microwave ECR nitrogen plasma forms the AlN tunnel barrier inside the sputtering system. The process reproducibly yields highly transparent junctions with $R_nA$ down to $\sim 3~\Omega\cdot\mu\mathrm{m}^2$ while maintaining high subgap quality, $R_j/R_n \gtrsim 25$. Structural analysis (HRTEM/EELS) confirms the formation of a continuous $\sim 1.2$–$1.5$~nm AlN barrier at these transparencies. At the device level we established a consistent specific‑capacitance calibration from cryogenic $S$‑parameter extraction and on‑wafer probe data, showing that the specific capacitance of our AlN‑barrier SIS junctions is $\approx 25\%$ lower than that of our AlO$_x$ reference process. The lower $C_s$ reduces shunt reactance and relaxes RF matching constraints. We also tracked $R_n$ and $R_j/R_n$ through room‑temperature aging and annealing up to $200^{\circ}\mathrm{C}$; only modest changes in $R_n$ and essentially constant $R_j/R_n$ provide comfortable thermal headroom for back‑end processing and packaging.

At the circuit level, we validated the $C_s(R_nA)$ calibration in a re‑tuned copy of a well‑characterized 275–370~GHz mixer. Simply rescaling dielectric thicknesses to compensate the $\sim 25\%$ lower $C_s$ was sufficient to preserve, and in parts widen, RF/IF tuning, while achieving competitive DSB noise across the band. The measured performance is consistent with the independently extracted $C_s$ and with the improved subgap quality of the AlN junctions. These results show that aluminium nitride barriers can be advantageously used for SIS mixer designs and offer a design margin on impedance matching.

Beyond SIS mixer devices, the materials and process demonstrated here are directly relevant to high‑density superconducting electronics. This process is compatible with modern planarized Nb VLSI flows, like, e.g., \cite{tolpygo_planarized_2019}. Our ECR-plasma nitridation barrier‑formation method provides an alternative to conventional Nb/Al/AlO$_x$/Nb technology in such flows, especially where higher $J_c$ or lower $R_nA$ is desired without sacrificing subgap quality.

Finally, the demonstrated ability of ECR-plasma nitridation to grow high-quality, low-leakage and thermally stable nitride tunnel barriers could be seen as an attractive tunnel barrier growth alternative in other material systems. For example, in emerging Ta-based Josephson junction circuits, e.g., \cite{bland_millisecond_2025}, ECR-plasma nitridation of Ta electrodes may become an alternative to oxidation or direct TaN deposition \cite{bhatia2025TaN}.

\section{\label{bibliography}Bibliography}
\bibliography{_APL_AlN-barrier_SIS_junctions.bib}

\end{document}